\documentclass[twocolumn,showpacs,superscriptaddress,prl]{revtex4}%,eqsecnum
\usepackage{xspace,amsfonts,amssymb,graphicx}
\begin{document}
\title{The entanglement of indistinguishable particles
shared between two parties}
\author{H.M. Wiseman}
\email{H.Wiseman@griffith.edu.au}
\affiliation{Centre for Quantum Computer Technology, Centre for Quantum Dynamics, School of Science, Griffith University,  Brisbane,
Queensland 4111 Australia. }%
\author{John A. Vaccaro}
\affiliation{Centre for Quantum Computer Technology, Centre for Quantum Dynamics, School of Science, Griffith University,  Brisbane,
Queensland 4111 Australia. }%
\affiliation{Division of Physics and Astronomy, University of
Hertfordshire, Hatfield AL10 9AB, UK.}%Queensland

\newcommand{\beq}{\begin{equation}}
\newcommand{\eeq}{\end{equation}}
\newcommand{\bqa}{\begin{eqnarray}}
\newcommand{\eqa}{\end{eqnarray}}
\newcommand{\nn}{\nonumber}
\newcommand{\nl}[1]{\nn \\ && {#1}\,}
\newcommand{\erf}[1]{Eq.~(\ref{#1})}
\newcommand{\erfs}[2]{Eqs.~(\ref{#1})--(\ref{#2})}
\newcommand{\dg}{^\dagger}
\newcommand{\rt}[1]{\sqrt{#1}\,}
\newcommand{\smallfrac}[2]{\mbox{$\frac{#1}{#2}$}}
\newcommand{\half}{\smallfrac{1}{2}}
\newcommand{\bra}[1]{\langle{#1}|}
\newcommand{\ket}[1]{|{#1}\rangle}
\newcommand{\ip}[2]{\langle{#1}|{#2}\rangle}
\newcommand{\sch}{Schr\"odinger }
\newcommand{\schs}{Schr\"odinger's }
\newcommand{\hei}{Heisenberg }
\newcommand{\heis}{Heisenberg's }
\newcommand{\bl}{{\bigl(}}
\newcommand{\br}{{\bigr)}}
\newcommand{\ito}{It\^o }
\newcommand{\str}{Stratonovich }
\newcommand{\dbd}[1]{{\partial}/{\partial {#1}}}
\newcommand{\sq}[1]{\left[ {#1} \right]}
\newcommand{\cu}[1]{\left\{ {#1} \right\}}
\newcommand{\ro}[1]{\left( {#1} \right)}
\newcommand{\an}[1]{\left\langle{#1}\right\rangle}
\newcommand{\st}[1]{\left|{#1}\right|}
\newcommand{\implies}{\Longrightarrow}
\newcommand{\tr}[1]{{\rm Tr}\sq{ {#1} }}
\newcommand{\del}{\nabla}
\newcommand{\du}{\partial}
\newcommand{\singlecol}{\end{multicols}
     \vspace{-0.5cm}\noindent\rule{0.5\textwidth}{0.4pt}\rule{0.4pt}
     {\baselineskip}\widetext }
\newcommand{\doublecol}{\noindent\hspace{0.5\textwidth}
     \rule{0.4pt}{\baselineskip}\rule[\baselineskip]
     {0.5\textwidth}{0.4pt}\vspace{-0.5cm}\begin{multicols}{2}\noindent}
\renewcommand{\section}[1]{{\em #1}.}
\renewcommand{\subsection}[1]{}

\begin{abstract}
Using an operational definition we quantify the entanglement, $E_{\rm P}$,
between two parties who share an arbitrary pure state of $N$
indistinguishable particles. We show that $E_{\rm P}  \leq
E_{\rm M}$,  where $E_{\rm M}$ is the bipartite entanglement
calculated from the
mode-occupation representation. Unlike $E_{\rm M}$, $E_{\rm P}$ is
{\em super-additive}. For example, $E_{\rm P}=0$ for any
single-particle state, but
 the state $\ket{1}\ket{1}$, where both modes are split between the two parties,
has $E_{\rm P} = 1/2$. We discuss how this relates to quantum
correlations between particles, for both fermions and bosons.
\end{abstract}


\maketitle

%\section{Introduction}
Entanglement lies at the heart of quantum
mechanics, and is profoundly important in quantum information (QI) \cite{NieChu00}.
It might be thought that there is nothing new to be said about bipartite
entanglement if the shared state $\ket{\Psi_{AB}}$ is pure. In ebits,
the entanglement is simply \cite{HorHorHor00}
\beq \label{defent}
E(\ket{\Psi_{AB}}) =
S(\rho_{A}).
\eeq
Here $S(\rho)$ is the binary von Neumann entropy $-\tr{\rho
\log_{2}\rho}$, and (since we will use unnormalized kets)
\beq
\rho_{A} = {\rm
Tr}_{B}[\ket{\Psi_{AB}}\bra{\Psi_{AB}}]/\ip{\Psi_{AB}}{\Psi_{AB}}.
\eeq
However, in the context of indistinguishable particles, a little
consideration reveals a less than clear situation, which has been the
subject of recent controversy \cite{Zan02,Git02,PasYou01,Sch01,Li01}.

Consider, for example, a single particle in an equal superposition
of being with Alice and with Bob. In the mode-occupation, or Fock,
representation, the state is
\beq
\ket{0,1}+\ket{1,0}.
\eeq
Here we are following the conventions of writing Alice's
occupation number(s) followed by Bob's, separated by a comma, and
of omitting any modes that are unoccupied. On the face of it, this
is an entangled state with one ebit, and such a state has been
argued to show nonlocality of a single photon
\cite{TanWalCol91,Har94}. However, the particle's wavefunction, in
the co-ordinate representation, is of the form
\beq
\psi_{A}(x) + \psi_{B}(x),
\eeq
where the subscripts indicate where the wavepackets are localized in
co-ordinate ($x$) space. In this representation,
the above entanglement is not apparent, and indeed it has been argued
that nonlocality cannot be a single-particle effect
\cite{GreHorZei95} (although see Ref.~\cite{BjoJonSan01}).

As a second example, consider a two-particle state where Alice has
one particle and Bob the other. In the mode-occupation picture,
the state is $\ket{1,1}$, which appears unentangled. But since
these are identical particles, the wavefunction must be
symmetrized as
\beq
\psi_{A}(x)\psi_{B}(y) \pm \psi_{A}(y)\psi_{B}(x) ,
\eeq
for bosons and fermions respectively. This has the appearance of an
entangled state.

Finally, consider another two-particle state, but this time where
the two particles are prepared and shared as in the first example,
but in different modes. In the Fock representation, this state is
entangled:
\bqa
&& (\ket{0,1}+\ket{1,0})(\ket{0,1}+\ket{1,0})
\nl{=} \ket{00,11} + \ket{01,10}
+ \ket{10,01} + \ket{11,00} .\label{halfent}
\eqa
The corresponding wavefunction
\bqa
&&[\psi_{A1}(x) + \psi_{B1}(x)][\psi_{A2}(y) + \psi_{B2}(y)]
\nl{\pm} [\psi_{A1}(y) + \psi_{B1}(y)][\psi_{A2}(x) + \psi_{B2}(x)]
\eqa
also has the appearance of an entangled state.

In this Letter we give an {\em operational} definition of
entanglement between two parties who share an arbitrary pure state
of indistinguishable particles. Applying this to the three states 
introduced above yields an entanglement (in ebits) of $0$, $0$, and $1/2$ respectively.

To justify these (non-obvious)
answers, we proceed as follows. First we 
{\em define} precisely what we mean 
when we use the term {\em particle}. Next  we review two
{\em previous measures} of entanglement. The first, as championed in
Refs~\cite{Zan02,Git02} for example,
we call entanglement of modes. The second, following Pa\v{s}kauskas
and You (PY) \cite{PasYou01} and others, we call the quantum correlation
between
two particles. Then we introduce our own concept, the {\em entanglement
of particles}. We show that (at least for bosons) our criterion for
entanglement is
 stronger than both previous criteria, and even their conjunction.
 We illustrate our measure with
 several {\em examples}, and finally prove a number of its {\em properties} 
including super-additivity.

\section{Definition of ``particle''} For the purpose of our analysis, 
a particle is a discrete entity which is indistinguishable from 
other particles of the same type, and for which conservation laws 
imply a particle number superselection rule (SSR).
That is, the creation of superpositions of different number
eigenstates by any means (including measurements) is
forbidden. Examples of such ``genuine'' particles are: electrons; 
and Hydrogen atoms in a particular electronic state.
Note that the conservation law need not be for the
particle number itself. For example, composite particles such as
atoms can be constructed from their constituents and so their
number is not conserved. Nevertheless there is still a SSR because
of the fundamental conservation laws of baryon and lepton numbers
of the constituents \cite{Sud86} and this is all we need here.
However, it is usual in QI processing for the
number of particles to be conserved \cite{VerCir03}.  Thus, without loss
of generality, we will assume a conservation law for particle
number. Note also that such a law does not apply to the quanta of
excitation (which are always bosonic) of oscillators which can be
excited classically, such as photons or excitons.

\section{Previous Measures}
\subsection{Entanglement of Modes}
One measure of entanglement of identical particles is what we
call the entanglement of modes $E_{\rm M}$. This  is simply determined by
calculating \erf{defent} for the bipartite state in the Fock
representation. Since we are concerned only with genuine particles,
we will henceforth
assume that the
joint state $\ket{\Psi_{AB}}$ contains exactly $N$ particles.
Clearly $E_{\rm M}$ is
independent of whether the particles are bosons or fermions, but
in the latter case the occupation numbers are limited to 0
or 1.

\subsection{Quantum correlation between particles}
A completely different concept can be considered for the case where $N=2$,
namely whether one {\em particle} is entangled with the other
\cite{PasYou01,Sch01,Li01}.
This is conceptually quite different from both
the entanglement between two spatially separated parties, and the
entanglement between two distinguishable particles (which in principle
could be separated, unlike identical particles). To emphasize these
differences, PY \cite{PasYou01} follow Schliemann {\em et al.}
\cite{Sch01} in referring
instead to {\em quantum correlations} (QC) between particles.
Since there is no notion of spatial separation, we
drop for now the $AB$ subscript on the state $\ket{\Psi}$.

The QC as defined by PY
is different for bosons and fermions. The QC
between bosons is given in bits as
\beq
S_{b} = S(\rho^{(1)}),
\eeq
where $S$ is the binary entropy as above, and where $\rho^{(1)}$ is
the single-particle mixed state. In some particular mode basis, the state
matrix is
\beq
[\rho^{(1)}]_{\mu'\mu} = {\bra{\Psi}c\dg_{\mu}
c_{\mu'}\ket{\Psi}}/\sum_{\nu}{\bra{\Psi}c\dg_{\nu}
c_{\nu}\ket{\Psi}},
\eeq
where $c_{\mu}$ is the boson annihilation operator for mode $\mu$.
%We are using the convention that
%the tilde indicates an unnormalized state matrix, which must be
%divided by its own trace to yield the normalized state matrix
%$[\rho^{(1)}]_{\mu\mu'}$.
It turns out that for bosons it is always possible to find a basis
where the two-particle state can be written as
\beq
\ket{\Psi} = \sum_{\mu} \beta_{\mu} \frac{1}{\sqrt{2}}
(c_{\mu}\dg)^{2}\ket{\mathbf 0},
\eeq
where $\ket{\mathbf 0}$ is the state containing no particles, and
where the $1/\sqrt{2}$ corrects a typographical error in Ref.~\cite{PasYou01}.
From this it
can be shown that \cite{PasYou01}
$
S_{b} = H(\{|\beta_{\mu}|^{2}\}_{\mu})$,
where here $H$ is the binary {\em
Shannon} entropy \cite{NieChu00} for the probability  distribution
$\{|\beta_{\mu}|^{2}\}_{\mu}$.
Only the state $\ket{2}$,
with both bosons in one
mode, is uncorrelated.

For fermions, the single-particle mixed state and its entropy
are defined in precisely
the same way, but where the $c_{\mu}$s are fermion annihilation operators.
However PY say the
QC between fermions is, in bits,
$
S_{f} - 1
$.
This curious difference from the bosonic case is motivated as follows.
For fermions it is always possible to find a basis
where the two-particle state can be written as
\beq
\ket{\Psi} = \sum_{\nu} \phi_{\nu}  c_{2\nu}\dg c\dg_{2\nu-1}\ket{\mathbf 0}.
\eeq
From this it
can be shown that
$
S_{f} - 1 = H(\{|\phi_{\nu}|^{2}\}_{\nu})$.
The least-correlated state is a state of the form $\ket{1}
\ket{1}$, with
one fermion in each of two modes. This state has an entropy of 1, but
a QC of 0 according to PY.
%Note that if a state in a given Fock
%representation is a valid state for fermions (i.e. all occupation
%numbers are 0 or 1) then it will also be a valid state for bosons
%and $S_{b}=S_{f}$.
Thus one has the curious situation that the ``same''
state, such as $\ket{1}\ket{1}$ would be considered quantum correlated for
bosons but uncorrelated for fermions
% \footnote{Li {\em et al.} \cite{Li01}
%do suggest that the quantum
%correlation should be calculated in the same way for bosons and
%fermions, but choose to keep the fermion definition unchanged from
%PY. This has the consequence that   the boson state $\ket{2}$, or
%$\psi(x)\psi(y)$,  is considered
%``entangled''!}. five lines.

\section{Entanglement of Particles}
We wish to define the entanglement $E_{\rm P}$ between two distant parties,
Alice and Bob,  who
share some state of $N$ indistinguishable particles. An obvious
question is, what is wrong with $E_{\rm M}$ as
defined above? The answer is that it fails to take into account the
SSR for particle number. To fully use the supposed entanglement
$E_{\rm M}$ they share,
Alice and Bob in general
must be able to arbitrarily measure and manipulate their
local systems. Unless Alice's (and hence Bob's) state
happens to be a state of %a mixture of  states of WAS WRONG
definite particle number, this will mean violating
the SSR for particle number. For example,
the teleportation protocol
in Ref.~\cite{Git02} relies upon such forbidden operations.
Thus $E_{\rm M}$
 in general over-estimates the available entanglement  $E_{\rm P}$.

To be specific, say that in addition to all of the indistinguishable
particles which Alice and Bob may use in the experiment, their
quantum state $\ket{\Psi_{AB}}$ includes a conventional quantum register each,
initially in a product state. The {\em operational
definition} of $E_{\rm P}$ is the
maximal amount of
entanglement which Alice and Bob can produce between their quantum
registers by local operations (LOs) \footnote{Classical communication makes no difference 
here.}. 
%\footnote{It turns out that CC (which
%clearly must not include the transfer of the particles in
%question) is not necessary.}. two lines
Since the registers of Alice and
Bob consist of distinguishable
qubits, this entanglement can be computed by the
standard measure. As a consequence of
the particle number SSR, this entanglement will be
given not by the mode entanglement of $\ket{\Psi_{AB}}$, but by
\beq \label{ErhoAB}
   E_{\rm P}(\ket{\Psi_{AB}}) \equiv
   \sum_n P_n E_{\rm M}(\ket{\Psi_{AB}^{(n)}})\ .
\eeq
Here $\ket{\Psi_{AB}^{(n)}}$ is $\ket{\Psi_{AB}}$ projected
into the subspace of fixed {\em local} 
particle number ($n$ for Alice, $N-n$ for Bob),
\beq
   \ket{\Psi_{AB}^{(n)}} = \Pi_{n}\ket{\Psi_{AB}}\ ,
   \label{proj}
\eeq
%$\Pi_{n}$ is the projector for the subspace where Alice's particle
%number $\hat{n}$ has the value $n$ (and so Bob's is $N-n$),
and $P_n$ is the probability $\ip{\Psi_{AB}^{(n)}}{\Psi_{AB}^{(n)}}/\ip{\Psi_{AB}}{\Psi_{AB}}$.

To see this explicitly, say Alice and Bob perform the optimal LOs 
to change $\ket{\Psi_{AB}}$ into $\ket{\Psi'_{AB}}$ in
which their registers have entanglement $E_{\rm P}$. Alice could
now measure her local particle number $\hat{n}$, and this would
not affect $E_{\rm P}$ on average. But since LOs conserve
local particle number this would be true even if Alice were to
measure $\hat{n}$ {\em before} applying the LOs. %\cite{meas_n}. 
This measurement would collapse the state
$\ket{\Psi_{AB}}$ into the state $\ket{\Psi_{AB}^{(n)}}$, with
probability $P_n=\ip{\Psi_{AB}^{(n)}}{\Psi_{AB}^{(n)}}$, where $n$
is the measurement result. Now since this is a state of definite
local particle number for both parties, there are no conservation
laws that prevent local unitaries from transferring all of its
entanglement, $E_{\rm M}(\ket{\Psi_{AB}^{(n)}})$
%[which equals $E_{\rm M} (\ket{\Psi_{AB}^{(n)}})$ in this case],
to the quantum registers.
To obtain $E_{\rm P}$ as defined above one simply averages over
the result $n$, yielding \erf{ErhoAB}.

\subsection{Relation to Previous Concepts}
We can relate $E_{\rm P}$ to both previous concepts defined above.
First, since projecting onto the local particle number eigenspace is a local
operation, it can only decrease entanglement \cite{HorHorHor00}.
It follows that
\beq
E_{\rm P}(\ket{\Psi_{AB}}) \leq E_{\rm M}(\ket{\Psi_{AB}}),
\eeq
so that the entanglement of modes is necessary for the
entanglement of particles.
%Here we have generalized the definition
%(\ref{ErhoAB}) by allowing for an initially mixed state
%$\rho_{AB}$.

Second, it turns out (for bosons at least) that QC between particles is
also necessary for entanglement of particles.
%This is simple to see.
Recall that the QC was defined by YP only
for two particles. If the QC between two bosons
is zero then there is some choice of
modes such that they are in the same mode. In general
this mode will be split between an Alice mode and a Bob
mode, with coefficients $\alpha$ and $\beta$. Then in the Fock picture,
the two-boson state is
\beq
\frac{1}{\sqrt{2}}(\alpha a\dg + \beta b\dg)^{2}\ket{0,0} = \alpha^{2} \ket{2,0} +
\sqrt{2} \alpha\beta \ket{1,1} + \beta^{2}\ket{0,2}, 
\eeq
where $a$ and $b$ are the annihilation operators for the
relevant mode on Alice's and Bob's side respectively.
Since the three
terms here have different local particle number, $E_{\rm P}=0$.
Thus entanglement of particles ($E_{\rm P}>0$)
 implies QC between bosons.

For fermions, the situation is not so clear cut. Consider the two-particle
state (\ref{halfent}). Applying \erf{ErhoAB},
Alice and Bob share half an ebit through these two identical particles.
Using modes split between Alice and Bob (as in the preceding
paragraph), this state can be rewritten as $\ket{1}\ket{1}$. As a
bosonic state, this would be considered by PY as exhibiting
QC, but not so as a fermionic state.
As discussed, the justification for the latter categorization
is the desire to have the least-correlated two-fermion state have no
QC, like the least-correlated two-boson state. Our analysis shows
that what is more relevant to bipartite entanglement of particles is
that the one-particle entropy of this state, $S_{f}=S_{b}=1$, is nonzero.

This conclusion is strengthened in that the entanglement of
particles, as we have quantified it, reduces in the two-particle
case to a modified version of the single particle entropy $S_{b}$
or $S_{f}$, as defined by PY. It does {\em not} reduce to
$S_{f}-1$, as they define the QC between fermions to be. We would
expect the QC between the two particles to correspond to the
entanglement between Alice and Bob only if Alice has just one of
the particles. Therefore we modify the QC between particles of PY
by defining the single particle state matrix to be
\beq
[\rho^{(1)}_{A}]_{k'k} =  {\bra{\Psi_{AB}^{(1)}}a\dg_{k}
a_{k'}\ket{\Psi_{AB}^{(1)}}} / \sum_{l} {\bra{\Psi_{AB}^{(1)}}a\dg_{l}
a_{l}\ket{\Psi_{AB}^{(1)}}}. 
\eeq
Here the operators $a_{k}$ are those acting on Alice's modes only.
It is then easy to verify that the weighted single-particle
entropy $P_{1}S(\rho_{A}^{(1)})$ is identical to the entanglement
of particles as defined above. This is because
$[\rho^{(1)}_{A}]_{k'k}$ is the same state matrix as ${\rm
Tr}_{B}[\ket{\Psi_{AB}^{(1)}}\bra{\Psi_{AB}^{(1)}}]$ in the basis
$\ket{k}=a\dg_{k}\ket{\mathbf 0}$, and the contributions to
$E_{\rm P}$ from the $\ket{\Psi^{(0)}}$ and $\ket{\Psi^{(2)}}$
terms are zero.  This result holds for either fermions or bosons.

\section{Examples}
To illustrate our measure $E_{\rm P}$ we have tabulated it, as well
as $E_{\rm M}$,
and (where appropriate) $S_{b}$ and $S_{f}$,
for various states in Table~1.
A number of features are worth noting. First, as proven above,
$E_{\rm P}
\leq E_{\rm M}$, so that $E_{\rm P} > 0 \implies E_{\rm M} > 0$. Second, where
the single particle entropy $S$ is defined, it is identical for bosons
and fermions ($S_{b} = S_{f}$), and satisfies $E_{\rm P} < S$.
%We conjecture that this is
%always the case, and it
This is consistent with the result we %certainly
have proven that $E_{\rm P} > 0 \implies S >0$. Third, even if
$E_{\rm M} > 0$ and $S > 0$, this does not imply that $E_{\rm P} >
0$, so our concept cannot be derived from these previous concepts.

\begin{table}[th]
    \caption{Entanglement or related measures
    (in bits) for various states under various measures.
    A ``$-$'' indicates that the measure is inapplicable to that state.
    $E_{\rm M}$
    is the bipartite entanglement of modes,
    $S_{b}$ ($S_{f}$) the single-particle entropies for a
    two-particle system of bosons (fermions), and $E_{\rm P}$
     the bipartite entanglement
    of particles proposed here. All states are given in the Fock
    representation, with a comma separating the occupation numbers for Alice's
    modes from those of Bob.}
    \label{table:entanglement}
    \centering
    \begin{ruledtabular}
    \begin{tabular}{|l|c|c|c|c|}%{lccccccc}
    \hline
    State & $E_{\rm M}$ & $S_{b}$ & $S_{f}$ & $E_{\rm P}$  \\
    \hline\hline
    $\ket{0,1}+\ket{1,0}$ & 1 & $-$ & $-$ & 0   \\
    \hline
    $\ket{1,1}$ & 0 & 1 & 1 & 0 \\
    \hline
    $(\ket{0,1}+\ket{1,0})(\ket{0,1}+\ket{1,0})$ & 2 & 1 & 1 & 1/2
    \\  \hline
    $\ket{0,2}+\ket{2,0}$ &  1 & 1 & $-$ & 0\\
    \hline
    $\ket{0,2}+\sqrt{2}\ket{1,1}+\ket{2,0}$ &  3/2 & 0 & $-$ & 0    \\
    \hline
    $\ket{01,10}+\ket{10,01}$ & 1 & 2 & 2 & 1 \\
    \hline
    $\ket{11,00}+\ket{00,11}$ & 1 & 2 & 2 & 0 \\
    \hline
    $(\ket{0,1}+\ket{1,0})^{\otimes N}$ & $N$ & $-$ & $-$ & $\sim N$ \\
    \hline
\end{tabular}
\end{ruledtabular}
\end{table}

All of the numbers in Table~1 are trivial to calculate from the above
definitions, except for the asymptotic result. This is the
entanglement between Alice and Bob if  $N$
indistinguishable particles are prepared independently in $N$ different
modes, each of which is split equally between Alice and Bob. For large $N$, 
standard statistical mechanics
arguments \cite{KitKro80}  yields
\bqa
E_{\rm P}[(\ket{0,1}+\ket{1,0})^{\otimes N}]
&=& 2^{-N} \sum_{n=0}^{N}{N\choose n}\log_{2}{N\choose n} \nn \\
&\sim&  N - \frac{1}{2}\log_{2}(N) - \delta. \label{asym1}
\eqa
where $\delta = (-1 + \log_{2}\pi + 1/\ln 2)/2 \approx 1.047096$. %one line

\section{Properties} The astute reader will have noticed
that our measure of entanglement fails to satisfy the postulate of
partial additivity identified in Ref.~\cite{HorHorHor00}. That is,
\beq
E_{\rm P}(\ket{\Psi}^{\otimes C}) \neq C E_{\rm P}(\ket{\Psi}).
\eeq
The reason is that the states $\ket{\Psi}$ in the tensor product
are not truly independent of each other due to the
indistinguishability of the particles. If the subsequent terms in
the tensor product represented states of different species of
particle, the entanglement would be additive.

In fact, for arbitrary pure states of indistinguishable
particles, the entanglement is {\em super-additive}:
\beq \label{superad}
E_{\rm P}(\ket{\Psi}\otimes\ket{\Phi}) \geq
E_{\rm P}(\ket{\Psi})+E_{\rm P}(\ket{\Phi}) .
\eeq
This can be seen as follows.
%First,
%\bqa \label{defforPsi}
%E_{\rm P}(\ket{\Psi}) &=& \sum_{n=0}^{N}\ip{\Psi^{(n)}}{\Psi^{(n)}}S(\rho^{(n)}_{A}),
% \\ \label{defforPhi}
%E_{\rm P}(\ket{\Phi}) &=& \sum_{m=0}^{M}\ip{\Phi^{(m)}}{\Phi^{(m)}}S(\sigma^{(n)}_{A}).
%\eqa
%Here we are
%using $\tilde\rho_{A}^{(n)}$ and $\sigma_{A}^{(m)}$ for Alice's
%reduced state for $\ket{\Psi^{(n)}}$, and $\ket{\Phi^{(m)}}$
%respectively.
Say $\ket{\Psi}$ and $\ket{\Phi}$ are states with $N$ and $M$
particles respectively. We use $\rho_{A}^{(n)}$ and
$\sigma_{A}^{(m)}$ for the reduced state of $\ket{\Psi^{(n)}}$
and $\ket{\Phi^{(m)}}$ respectively, and  $p$ for the total
particle number, on Alice's side. Then, defining
$w_{nmp}=\delta_{p,n+m}
\ip{\Psi^{(n)}}{\Psi^{(n)}}\ip{\Phi^{(m)}}{\Phi^{(m)}}$,
\erf{ErhoAB} implies that $E_{\rm P}(\ket{\Psi}\otimes\ket{\Phi})$
equals
\beq
\sum_{p=0}^{N+M}
S\ro{\frac{\sum_{n,m}w_{nmp}\,\rho_{A}^{(n)}\otimes\sigma_{A}^{(m)}}
{\sum_{n,m}w_{nmp}}} \sum_{n,m}w_{nmp} .
\eeq
Using the concavity of the entropy \cite{NieChu00}, we obtain
\bqa
E_{\rm P} &\geq& \sum_{p=0}^{N+M}
\sum_{n,m}w_{nmp}
S(\rho_{A}^{(n)}\otimes\sigma_{A}^{(m)}) \label{concave} \\
&=& \sum_{p=0}^{N+M} \sum_{n,m} w_{nmp}
[S(\rho_{A}^{(n)})+S(\sigma_{A}^{(m)})].
\eqa
Applying \erf{ErhoAB} to $\ket{\Psi}$ and $\ket{\Phi}$, this
expression reduces to the right-hand side of \erf{superad}.

Now the inequality in \erf{concave} is an equality iff (if and only if) all
states in the sum are identical \cite{NieChu00}.
But actually they are orthogonal, so
the equality in \erf{superad} is satisfied iff   there is
only one element in the sum. That is, iff at least one of $V_{\Psi}$ or
$V_{\Phi}$ is $0$. Here $V_{\Theta}$ is the variance
in the number of particles on Alice's side for state $\ket{\Theta}$.

The above results also suggest that with a large number of
copies $C$, the mode entanglement is recovered:
\beq \label{asym2}
\lim_{C\to\infty} {E_{\rm P}(\ket{\Psi}^{\otimes C})}/
{E_{\rm M}(\ket{\Psi}^{\otimes C})} = 1 .
\eeq
 This can be established using the same techniques \cite{KitKro80} that gave
\erf{asym1}.
From the central limit theorem, the number of significant
terms in $\ket{\Psi}^{\otimes C}$ with
different particle number $n$ on Alice's side is of order $\sqrt{C
V_{\Psi}}$. For this state,
the entropy of a typical $\rho_{A}^{(n)}$  is therefore of order
$\log_{2}(\sqrt{C V_{\Psi}})$ smaller than the entropy of $\rho_{A}$.
That is,
\beq
E_{\rm P}(\ket{\Psi}^{\otimes C}) \sim
C E_{\rm M}(\ket{\Psi}) - \frac{1}{2}\log_{2}(V_{\Psi}C) + O(1),
\eeq
from which \erf{asym2} follows.

%\section{Conclusion}
In conclusion, from an {\em operational definition} we have
quantified $E_{\rm P}$, the entanglement of $N$ indistinguishable
particles shared between two parties. Our criterion for
entanglement of particles ($E_{\rm P}>0$) is stronger (in a
mathematical sense) even than the conjunction of two previous
concepts in this area. These are: the entanglement of modes
($E_{\rm M}>0$); and, for the two-particle case, correlations
between the particles ($S>0$, where $S$ is the single-particle
entropy). For asymptotically many copies of the state, $E_{\rm P}
\to E_{\rm M}$. However, unlike $E_{\rm M}$, $E_{\rm P}$ is {\em
super-additive}. This unusual characteristic of $E_{\rm P}$
reflects the indistinguishability of the particles.  It has
implications for QI processing and Bell-type
nonlocality, and requires further investigation (see, for comparison, Ref.~\cite{VerCir03}).  The
generalization of our work to multipartite entanglement 
%(with the case of {\em N parties} being of particular interest) 
is another
area for future exploration.

%\begin{acknowledgments}
\vspace{1ex}
We are grateful for discussions with J. Ruostekoski,
T. Osborne, M. Nielsen,
J. Dodd, and M. Bremner.
This work was supported by the Australian Research Council.
%\end{acknowledgments}

\end{document}